\documentclass[11pt]{amsart}
\usepackage[utf8]{inputenc}
\usepackage{amsmath}		
\usepackage{amsthm}
\usepackage{algorithm}
\usepackage{algpseudocode}

\usepackage{amssymb,amsmath,amsfonts,amsthm,cite,mathrsfs}
\usepackage{wasysym, harmony} 
\usepackage{marvosym} 
\usepackage{enumitem}   
\usepackage{verbatim}
\usepackage{xcolor}
\usepackage[pdftex,colorlinks]{hyperref}

\usepackage{dsfont} 
\usepackage{graphicx}
\usepackage{float} 
\usepackage{listings} 
\usepackage[normalem]{ulem}
\usepackage{geometry}[margin=1in]

\usepackage{algorithmicx}
\usepackage{subfig}

\definecolor{pink}{rgb}{1,0,1}
\newcommand{\R}{\mathbb{R}}

\theoremstyle{remark}

\theoremstyle{definition}
\newtheorem{defn}{Definition}

\title{Diversity is Key:  Fantasy football dream teams under budget constraints}
\title{Diversity is Key: Fantasy football dream teams under budget constraints} 
\author{Josef Gullholm}
\address{Department of Mathematical Sciences \\Chalmers University of Technology and The University of Gothenburg\\SE-41296, Gothenburg}

\author{Jil Kl\"under} 
\address{Institute of Practical Computer Science \\ Leibniz Universit\"at Hannover \\  Welfengarten 1 D-30167 Hannover} \email{jil.kluender@inf.uni-hannover.de}

\author{Julie  Rowlett}
\address{Department of Mathematical Sciences \\Chalmers University of Technology and The University of Gothenburg\\SE-41296, Gothenburg}

\email{julie.rowlett@chalmers.se}
\author{Jonathan St\aa lberg}
\address{Department of Mathematical Sciences \\Chalmers University of Technology and The University of Gothenburg\\SE-41296, Gothenburg}

\begin{document}
\maketitle

%Scope statement:  In numerous contexts, one seeks to assemble the highest performing team subject to a budget constraint. Rowlett et. al. developed a game theoretic model for competing teams that predicts features of successful teams.  Here, we analyzed the characteristics of top performing fantasy football teams to compare with the game-theoretic predictions. For several seasons and several budgets we determined the highest scoring fantasy football teams.  Without a sophisticated data-reduction, this would take 200-300 thousand years on a standard computer.  To analyze the data we therefore first created a powerful data cleaning algorithm that could be generalized to numerous other contexts and can therefore be of independent interest, as it increased  the efficiency by one quintillion percent.  After determining the composition of the highest scoring team within each budget, we analyzed all the teams. We considered several variables that are essential to football and discovered that a common feature of these variables’ distributions across the players is diversity.  We calculated that such diversity would not be expected from a randomly selected team.  Our analysis is consistent with the predictions of Rowlett et. al's game theoretic model and may therefore have broad implications contexts involving high performing teams.     

\begin{abstract}
Given a fixed budget for player salaries, what is the distribution of salaries of the top scoring teams? We investigated this question using the wealth of data available from fantasy premier league football (soccer).  Using the players' data from past seasons, for several seasons and several different budget constraints, we identified the highest scoring fantasy team for each season subject to each budget constraint.  We then investigated quantifiable characteristics of these teams.  Interestingly, across nearly every variable that is significant to the game of football and the budget, these top teams display diversity across these variables.  Furthermore, randomly assembled teams would statistically \em not \em display such diversity across these variables.  Our results indicate that diversity across these variables, including salaries, is a general feature of top performing teams.  Moreover, in the process of obtaining these results we developed a data cleaning (or data reduction) algorithm that drastically reduced the amount of data to be analyzed.  
\end{abstract}

\section{Introduction}
Sports have fascinated and entertained people for thousands of years \cite{Thuillier2004}.  People enjoy not only playing sports but also being a fan and cheering for their favorite competitor or team.  Fans also like to place wagers on sporting events and competitions \cite{Holt2011}.  Sports betting has a long history dating back thousands of years, like in ancient Rome where bets could be placed at chariot races \cite{Rodolfo1892}. Betting on sports continued through the years into modern times, and today sports betting is a major industry attracting billions of customers \cite{Forbes}.  

While one can bet on real-life matches, fantasy sports allow fans to play the role of team manager and bet on the fantasy teams they create. Fantasy sports are a class of multiplayer games based on real life sport players’ performances.  They date back to at least the 1950s when Wilfred Winkenbach created fantasy golf \cite{Kissel2017, RuihleyVoice2021}.  Fantasy sports were introduced to academia in 1960 when Harvard University sociologist William Gamson started the ``Baseball Seminar'' where colleagues would form rosters that earned points based on the players' final standings in batting average, RBI, ERA, and wins \cite{Davis2006,gamson1976}.  The landmark ``Rotisserie League'' was also based on Major League Baseball and founded by Daniel Okrent in the 1980s \cite{Ploeg2021}.  Fantasy sports gained tremendous popularity during the 1990s as widespread internet access became available \cite{RuihleyVoice2021}, with the foundation of the Fantasy Sports Trade Association in 1998 \cite{BillingsFantasy2013}.  Over 50 million Americans play fantasy sports \cite{Foster2020, King2017, Dyreson2019}.  In India, there are around 100 million users playing Fantasy cricket \cite{Brettenny2012, Adhikari2020,Naha2021,Karthik2021,  Kaur2020}.  Globally, fantasy sports is a multibillion dollar industry \cite{South2019, Ruihley2021}.   

Anyone can play fantasy football for free.  For Fantasy Premier League (FPL), you begin with a budget of 100 million pounds and act as the manager of a Premier League football team.   Adhering to your budget, you select fifteen players: two goalkeepers, five defenders, five midfielders, and three forwards. The basic aim of FPL  is to accumulate as many points as possible across the season.  From your squad of 15 players, you must select a starting XI for each game week; a game week usually consists of 10 games played during the season.  The players you select for each game week receive points based on their real-world performances.  You can put your team in any formation, although there must always be one goalkeeper, at least three defenders and at least one forward selected at all times.  There are many further details involved in playing FPL which can be found on the official \href{https://www.premierleague.com/}{FPL homepage.}  

Unlike playing FPL, however, real-life team managers have \em different \em budgets for their teams.  It is therefore natural to ask, what is the best team one can create for a specific budget?  Using the available FPL data from past seasons, we investigate how many points all possible starting XI, subject to different budget constraints, score.  In this way, for each budget, we identify the best possible team one could create.  We reasonably expect the insights gained by the compositions of these teams may be useful and interesting in both real-world and fantasy sports.  Moreover, the methods we use to obtain our results may be of further interest, because a priori there is an impossibly large number of calculations required to analyze all possible teams that could be created within a given season.  For example, in season 2016--2017, there were 684 active players.   Considering all possible starting XI that could be formed with these players amounts to $684 \choose 11$ $ \approx 10^{23}$ combinations.  Calculating the scores of all of these teams could not be achieved in a human lifetime.

In spite of this seemingly insurmountable number of combinations, for 11 budget constraints, we determined for each budget constraint the best team according to total points earned by its players during a season.  We did this for 5 seasons starting from the 2016--2017 season and continuing to the 2020--2021 season.   We defined the lowest budget in order to compare teams comprised of players that actually had game time, because the absolute cheapest players often did not have any game time.  The highest budget was based on the best team one could build without imposing any budget constraint. This means that if we would have explored higher budgets we would not have obtained any further results.  Starting from the lowest budget and increasing incrementally to the highest budget resulted in a total of 11 budget constraints.   We then investigated the compositions and characteristics of the best teams for all 11 budgets for all 5 seasons.  Although we characterized the best teams cumulatively over an entire season, the same approach could be used iteratively to provide insights into which trades to make during a season. Moreover, our data reduction methods that allowed us to analyze a seemingly impossibly large data set could be of independent interest due to its theoretical basis and therewith lack of constraint to any single field of application.

\section{Methods:  data collection, reduction, and analysis} \label{s:data} 
We used data collected from five Fantasy Premier League (FPL) seasons, starting with the 2016--2017 (2017) season and continuing through the 2020--2021 (2021) season.  We gathered the most recent Premier League season was from the FPL websites’ API with permission from FPL. Since only the current season is available on the websites’ API, we obtained data for the previous seasons from a \href{https://github.com/vaastav/Fantasy-Premier-League/tree/master/data}{GitHub repository} \cite{Git} which has collected data from several FPL seasons. 

We used the data to calculate all players' summary statistics over a full season. The variables we used to construct the best teams were the \textit{cost} at the end of the season and the \textit{total points} gathered for the whole season. We also needed the position of each player, denoted \textit{element\_type}.  

\begin{defn} \label{def:element_type} For notational convenience, we use the numbers 1, 2, 3, 4 to represent each of the four positions: 
\begin{itemize}
    \item 1 - Goalkeeper
    \item 2 - Defender
    \item 3 - Midfielder 
    \item 4 - Forward.
\end{itemize} 
\end{defn} 

In the 2017 season  there were 684 players, and in 2021 there were 714 players.  We organized this information as shown in Table \ref{tab:example_rows}. 

\begin{table}[H]
\centering
    \begin{tabular}{l|l|l|l}
    Player\_id & Cost [£100k] & Total\_points & Element\_type \\ \hline
    28        & 105  & 139          & 4            \\
    29        & 73   & 90           & 2            \\
    30        & 43   & 73           & 1            \\
    31        & 43   & 0            & 3           
    \end{tabular}
    \caption{This is an example of four rows of data from a fantasy season. The cost is in units of 100 000 GBP at the end of the season.  The element\_type represent the position of the player:  Goalkeeper (1), Defender (2), Midfielder (3), and Forward (4).}
    \label{tab:example_rows}
\end{table}

\subsection{Data reduction} \label{sec:data_clean}
If we were to calculate \em all \em fantasy teams that could theoretically be created during a single season, this would require a few thousand years with our best algorithm on a standard computer.  To see this, for example in the 2017 season, there were 684 total players.  So, one would need to analyze ${684 \choose 11} \approx 10^{23} $
theoretical teams, since a team always has 11 players.  For other seasons, the order of magnitude is similarly large.  Of course, not all such theoretical teams are possible in fantasy football, because a team must have one of the allowed formations (goalkeeper-defenders-midfielders-forwards): 1-3-4-3, 1-3-5-2, 1-4-4-2, 1-4-5-1, 1-4-3-3, 1-5-3-2 and 1-5-4-1.  However, even imposing this restriction does not reduce the amount of calculations to a manageable quantity. We realized, however, that to determine the \em highest \em scoring team subject to a specific budget constraint, it is not necessary to consider \em all \em fantasy teams.  With this in mind, we created a data reduction algorithm that \em significantly \em reduced the number of teams in each budget bracket.  This type of algorithm could be used for other analyses that compare a certain metric of performance subject to a constraint on resources.  

The data reduction algorithm discards players that we can rigorously prove would never be a part of the best eleven for a specific budget constraint by determining when there are cheaper players of the same type (position) with higher total points.  The player with a \em higher \em cost and \em lower \em number of points would never be chosen as part of the best eleven if there are cheaper player(s) of the same type (position) with higher points.  Similarly, if one wishes to analyze the characteristics of the best performing collection of certain individuals or items according to a measure of performance subject to a budget constraint, then one could use an algorithm based on the one we present here.  In this sense, with our aim being to investigate the composition of the highest scoring fantasy team subject to specific budget constraints, it turns out that we can eliminate a huge amount of superfluous data.  Our data algorithm deletes the superfluous data corresponding to players and combinations that cannot be a part of the best eleven for each budget constraint.  This algorithm drastically reduces the number of calculations, reducing the runtime by several orders of magnitude, so that we can execute all calculations in a matter of hours on a standard computer.  Our data reduction algorithm on the players' data consists of the algorithm \ref{alg:clean} together with three steps in \S \ref{ss:3_steps} below.  

 \begin{algorithm}[H]
\begin{algorithmic}
\State $n \gets length(\text{position})$
\State $X \gets$ Data frame sorted by cost
\State $best \gets$ n first items in $X$
\For{$x \in X$}
\If{$x > min(best) $}
    \State Replace $min(best)$ with x
\Else
    \State Remove $x$ from X
\EndIf
\EndFor
\end{algorithmic}
\caption{This algorithm reduces the player data by removing that which will never be used in the best teams. When increasing the total cost we are able to pick more players with a higher cost, but this is only advantageous if those players also have higher points than the players with lower costs. This algorithm checks whether the team gets more total points if we select the new players that we can afford when increasing the budget. If we don't use the new players because they will not increase the total points, then we discard them.}
\label{alg:clean}
\end{algorithm}

\subsubsection{The three-step process} \label{ss:3_steps} 
We first compare for each cost value the total points obtained by each type of player (element types $1-4$).  Next we compare for each total point value the players' costs, for each element type.  Finally, we used Algorithm \ref{alg:clean} to determine if increasing the budget can result in higher points, or not.  These three steps are described more precisely below.  

\begin{enumerate}
 \item In the first step, for each cost value, we compare the total points for each element type (position).  For example, since there is always precisely one goalkeeper, for each cost value, we only keep one goalkeeper who, for that cost value has the highest points.  We use $n_e$ to denote the amount of players of element type (position) $e$ that it is possible to have in an allowed formation.   So, $n_1$ is always one, but for the other positions, this works a bit differently, because for example the number of defenders can be $3$, $4$, or $5$.  So, we allow $n_2$, which is the number of defenders (since element type 2 is defender) to be $3$, $4$, or $5$.  Then, for each cost $y$, we keep those $3$, $4$, or $5$ defenders (according to the specified value of $n_e$) who have the highest number of points, and we discard all the rest.   An example of this is shown in Table \ref{table:clean_ex1}.

 \item  In the second step, we consider for each number of points, the players' costs.  So, for each total point value $p$, for each element type $e$, for each of the possible values of $n_e$, we only keep the $n_e$ players with total points $p$ who also have the lowest costs.  An example of this is shown in Table \ref{table:clean_ex2}.   
 
 \item We deleted players that had \textit{less} points but a \textit{higher} cost according to Algorithm \ref{alg:clean}.  We started with the $n_e$ cheapest players for the given position $e$. We saved the total points for the $n_e$ highest scoring players that is affordable in a collection (i.e. under a certain budget).  After that, we increased the budget by 1, allowing the next player to be a part of the $n_e$ best players. If the new player had  higher total points than one of the previous players in the current $n_e$ best players, we put the new player in the $n_e$ best player collection and removed the cheapest one from that collection. We still kept the removed player since for a lower budget, it would be a part of the collection. If however, the new player wasn't included in the new collection it meant that, no matter the budget, this player would never be picked since there are cheaper players that generate more points. In that case we could delete that player. This process is shown in Table \ref{table:clean_ex3}. 
\end{enumerate}

\begin{table} 
\centering
\begin{tabular}{l|l}
Cost & Points   \\\hline
52   & 61   \\
52   & 30   \\
52   & 28   \\
52   & 2    \\
52   & 0    \\
52   & 0      \\
53   & 73   \\
53   & 10 
\end{tabular} 
\quad 
\begin{tabular}{l|l}
Cost & Points \\\hline
52   & 61     \\
52   & 30     \\
52   & 28     \\
\sout{52}   & \sout{2}\\
\sout{52}   & \sout{0}\\
\sout{52}   & \sout{0}      \\
53   & 73     \\
53   & 10    
\end{tabular}
\caption{This shows an example of the first step in our data reduction algorithm in the case that the total number of players of this element type can be three, like forwards.  The left shows the data before, and the right shows the data that is kept (and removed) in this step.  We keep the three highest scoring players at cost $52$ and discard all others who have cost $52$ and lower points than these top three.}
\label{table:clean_ex1} 
\end{table}

\begin{table}
\centering
\begin{tabular}{l|l}
Points & Cost \\\hline
70   & 44     \\
70   & 56     \\
70   & 56     \\
70   & 60\\
70 & 87\\
70   & 106\\
71   & 42     \\
73   & 51
\end{tabular}
\quad
\begin{tabular}{l|l}
Points & Cost \\\hline
70   & 44     \\
70   & 56     \\
70   & 56     \\
\sout{70}   & \sout{60}\\
\sout{70} & \sout{87}\\
\sout{70}   & \sout{106}\\
71   & 42     \\
73   & 51
\end{tabular}
\caption{This shows an example of the second step in our algorithm in the case $n_e=3$, so there are allowed formations with three players on a team of element type $e$.  The left shows the data before, and the right shows the data that is kept (and removed) in this step. We keep the three players that all have $70$ points as well as the lowest costs, and we discard the players that also have $70$ points but have \em higher \em costs.}
\label{table:clean_ex2} 
\end{table} 

\begin{table} 
\centering
\begin{tabular}{l|l}
Cost & Points \\\hline
\textbf{52}   & \textbf{61}    \\
\textbf{52}   & \textbf{30}     \\
\textbf{52}   & \textbf{28}     \\
53   & 73     \\
53   & 10   
\end{tabular}
\quad
\begin{tabular}{l|l}

Cost & Points \\\hline
\textbf{52}   & \textbf{61}    \\
\textbf{52}   & \textbf{30}     \\
52   & 28     \\
\textbf{53}   & \textbf{73}     \\
53   & 10   
\end{tabular}
\quad
\begin{tabular}{l|l}
Cost & Points \\\hline
\textbf{52}   & \textbf{61}    \\
\textbf{52}   & \textbf{30}     \\
52   & 28     \\
\textbf{53}   & \textbf{73}     \\
\sout{53}   & \sout{10}   
\end{tabular}
\caption{This shows the third step in our data reduction process.  The budget on the left is 156.  When we increase the budget to 157, we keep the players with cost 52 and have the highest points, and we replace the player with cost 52 and 28 points by the player with cost 53 and 73 points. Increasing the budget from 157 to 158 in the column on the right, we keep the same three players as for the budget 157, because the total points are already maximized. } 
\label{table:clean_ex3} 
\end{table} 

At this stage we have saved all players that are the best for each possible value of $n_e$ for that position corresponding to the possible team formations. After that we created all the combinations of the players in each group. For example, assume that we have 100 players in the group defenders which is element type $e=2$. Since every formation has at least $3$ defenders, we started by taking $n_2=3$, corresponding to those formations that have $3$ defenders.  We then calculated all the combinations possible for three players from that set of 100.  Next we did the same for all other possible values of $n_e$, which in the case of defenders means we calculated all the combinations possible for four players from the set of 100 as well as all the combinations possible for five players from the set of 100.  We did this for each of the four element types, $e$, and each of the possible values of $n_e$ for that element type.  When all the combinations for all positions were created, we then used Algorithm \ref{alg:clean} to get rid of position sets with a higher cost but lower points.  

We then sorted these position groups in ascending order of cost, so we started with the lowest cost.  So, for example, since $n_1$ the number of goalkeepers is always one, at each cost there is precisely one goalkeeper, namely the goalkeeper with the highest points for that cost.  For $n_2$, for each cost, we have $3$, $4$, or $5$ defenders for that cost, who are those with the highest points.  Starting with the lowest cost, we saved the points that the first combination had as a reference. If we increased the cost and did not receive higher points than the reference points, we discarded that combination. If we received higher total points, we set our reference as this new total point value and continued looping through all combinations. In the end, we combined the different position sets, for example 3,4 or 5 defenders with the goalkeeper, the midfielders and the forwards into the allowed formations  (goalkeeper-defenders-midfielders-forwards): 1-3-4-3, 1-3-5-2, 1-4-4-2, 1-4-5-1, 1-4-3-3, 1-5-3-2 and 1-5-4-1.  With these considerations, once we had assembled all the possible teams for all possible formations, we had approximately $7\times 10^7$ total teams remaining.  These algorithms therewith decreased the number of teams from on the order of $10^{23}$ to the order of $10^7$.  Our \href{https://github.com/gullholm/fantasy_master}{data} is available on github, as well as the \href{https://github.com/gullholm/fantasy_master/blob/main/cleaners.py}{data reduction algorithms} in Python. 

It is worth noting that there is a more elegant way to calculate the highest scoring team within each budget constraint by formulating an integer programming (IP) problem.  To do this we represent our data set with vectors in $\R^7$.  The first four components indicate the type of player.  A player of type $j$ has one in the $j^{th}$ component and $0$ in the three other of the first four components.  The fifth component of the vector is the cost of the player.  The sixth component of the vector is their total points, and the seventh component is their ID.  Then, one selects $11$ vectors from our (reduced) data set and computes their sum, denoted $s$.  The problem is then to maximize $s \cdot e_6$ subject to the constraints corresponding to the allowed formations and the budget:  $s \cdot e_1 = 1, \quad 3 \leq s \cdot e_2 \leq 5, \quad 3 \leq s \cdot e_3 \leq 5, \quad 1 \leq s\cdot e_4 \leq 3,$ 
$ s\cdot e_5 \leq \textrm{ budget constraint.} $ 
This problem can be solved in python with the following code.  The results of running this code, which we call lp-problem.py, are contained in our \href{https://github.com/gullholm/fantasy_master}{github repository}, and the results of this code agree with the results of our original algorithm.  

\begin{algorithm}[H]
\caption{Integer programming formulation in python for finding the best team within each budget constraint.}
\begin{small} 
\begin{verbatim} 
import pandas as pd
from pulp import LpProblem, LpVariable, lpSum, LpMaximize, value
def solve_optimization_problem(data, budget_constraint):
# Create a linear programming problem
prob = LpProblem("VectorSelection", LpMaximize)
# Decision variables
vectors = list(range(len(data)))
x = LpVariable.dicts("x", vectors, cat="Binary")
# Objective function
prob += lpSum(data[i][5] * x[i] for i in vectors), "Objective"
# Formation constraints
prob += lpSum(data[i][0] * x[i] for i in vectors) == 1, "GoalieConstraint"
prob += lpSum(data[i][1] * x[i] for i in vectors) >= 3, "DefenderConstraint"
prob += lpSum(data[i][1] * x[i] for i in vectors) <= 5, "DefenderConstraint2"
prob += lpSum(data[i][2] * x[i] for i in vectors) >= 3, "MidfielderConstraint"
prob += lpSum(data[i][2] * x[i] for i in vectors) <= 5, "MidfielderConstraint2"
prob += lpSum(data[i][3] * x[i] for i in vectors) >= 1, "FowardConstraint"
prob += lpSum(data[i][3] * x[i] for i in vectors) <= 3, "ForwardConstraint2"
# Budget constraint prob += lpSum(data[i][4] * x[i] for i in vectors) 
<= budget_constraint, "BudgetConstraint"
# Constraint to select exactly 11 vectors prob += lpSum(x[i] for i in vectors) 
== 11, "TotalPlayersConstraint"
prob.solve()

if prob.status == 1: # If the optimization problem is feasible
return int(value(prob.objective))
else: return None 
  
def get_players(filepath):
data = pd.read_csv(filepath, usecols=['element_type', 'now_cost', 
'total_points', 'id'])
players =[[1 if i == row['element_type'] - 1 else 0 for i in range(4)] 
+ [row['now_cost'],  row['total_points'], row['id']] 
for _, row in data.iterrows()]
return players

def calculate_optimal_team_scores(players):
obj_values = []  
for budget_constraint in range(500, 1001, 50):
obj_value = solve_optimization_problem(players, budget_constraint)
obj_values.append(obj_value)
return obj_values        

if __name__ == "__main__":
players = get_players('data/pl_csv/players_raw_2021.csv')
obj_values = calculate_optimal_team_scores(players)
print(obj_values)
\end{verbatim} 
\end{small} 
\end{algorithm} 

We note that this is a variation of the classic knapsack problem which consists of choosing an optimal subset of a set to carry in a knapsack of fixed size.  In this problem, optimal means the subset that fits within the knapsack and has the maximum value.  

\subsection{Following the money} \label{s:quantifying_diversity} 
Implementing the algorithm above, we determined the highest scoring fantasy teams that could be created subject to a given budget constraint.  We then analyzed different variables that are essential to football for these top performing teams subject to each budget constraint.  Perhaps the most important variable is \em money.  \em A team manager has a budget:  a limited amount of money that they can spend on the team.  How should they spend this budget on the players?  What are the salary distributions of the best performing teams?  Is the distribution similar to a normal distribution, with several players earning close to the average salary and a few outliers?  Or are there a few star players with very high salaries and several players with much lower salaries?  Or is there a wide range of salaries, dispersed somewhat evenly across the interval ranging from the lowest to the highest?  If this is the case, we would describe the distribution as \em diverse.  \em  We could ask the same question regarding the distribution of other variables that are essential to performance, like assists and goals.  How are the values of these variables across the players?  Are the values clumped like a normal distribution, or are they more evenly spread out across a range, like a \em diverse \em distribution?  

\begin{algorithm}[H]
\caption{Algorithm to determine if a team is diverse based on the bin approach.}\label{alg:div_disc}
\begin{algorithmic}
\State $Team\text{ }costs \gets$ List of the ind. cost of whole team 
\State $Containers \gets$ 11 container bins spanning the whole teams cost
\For{$Individual\text{ }cost$ in $Team\text{ }costs$}
\For{$\forall Bins \in Interval$}
\If{Individual cost $\in$ Bin}
\State $bin \gets 1$ 

\EndIf
\EndFor
\EndFor
\State $n \gets $\# empty container bins
\If{$n \leq 3 $}
    \State $Team \gets $ Diverse
\Else 
    \State $  Team \gets $ Not diverse
\EndIf

\end{algorithmic}
\end{algorithm}

In \cite{diversity}, Rowlett et. al. developed a game theoretic model that analyzes competition between teams of individuals subject to a resource constraint.  There they show that a diverse distribution of resources across the team's individuals is associated with team success in competition.  With this motivation, we introduced an intuitive `bin-method' to classify distributions as diverse or not.  For each team we calculated an eleven step interval grid.  At each step we made a subinterval, or bin, such that the union of all sub-intervals is the full range of the variable under consideration. For each team, we looped through all the individual players' values of the particular variable under consideration, and then we sorted them into the appropriate bins. If there were no empty bins, then we consider the distribution of that particular variable to be \em diverse, \em because it is analogous to the diverse distribution in the discrete model from \cite{diversity}.  However, since reality is not perfect or necessarily perfectly aligned with mathematical theorems like those of \cite{diversity}, here we still classify a distribution as diverse if a few of the bins are empty.   Thus, a distribution with at most $three$ empty bins is considered a diverse distribution.  

\section{Results:  the dream teams}  \label{s:results} 
We investigated the teams that performed the best within each budget constraint, and here present the analysis made on two seasons, 2017 and 2019, due to space limitations.  However, interested readers are referred to  \cite{mastersthesis} for further details of other seasons and further analyses.  It may also be interesting to note the total cost and total points for the best teams within each budget constraint in \ref{tab:best_2021_cost_and_points} for the 2021 season. 

\begin{table}[h]
\centering
\begin{tabular}{l|l|l}
Budget & Best total cost & Best total points \\\hline
500    & 500 & 1182      \\
550    & 550 & 1508      \\
600    & 600 & 1732      \\
650    & 648 & 1832      \\
700    & 700 & 1904      \\
750    & 746 & 1978      \\
800    & 797 & 2049      \\
850    & 849 & 2104      \\
900    & 900 & 2142      \\
950    & 940 & 2164      \\
1000   & 980 & 2178
\end{tabular}
\caption{The maximum budget is in the left column given in units 100 000 GBP.  Subject to this budget constraint for the total salaries, the middle column gives the actual cost of the best performing team.  The right column is the total points of that team.  This is shown for the season 2021.}
\label{tab:best_2021_cost_and_points}
\end{table}

When we analysed the total points for the best teams under each budget we obtained that the points increased logarithmically. As one would expect, the team with the highest points possible is that with the highest budget.  The name, cost, and points of all the players in the best team with the highest budget for the 2021 season is show in Table \ref{tab:best_2021_players}.

\begin{table}[h]
\centering
\begin{tabular}{l|c| c}
Name & Cost & Points \\\hline
Patrick Bamford & 66 & 194 \\
Jamie Vardy & 102 & 187\\
Harry Kane & 119 & 242\\
Heung-Min Son & 96 & 228\\
Marcus Rashford & 96 & 174\\
Bruno Miguel Borges Fernandes & 113 & 244 \\ 
Mohamed Salah & 129 & 231 \\
Stuart Dallas & 55 & 171 \\
Andrew Robertson & 73 & 161\\
Trent Alexander-Arnold & 78 & 160\\ 
Emiliano Martínez& 53 & 186\\\hline
Total & 980 & 2178 

\end{tabular}
\caption{This is the name, cost and points of all the players in the best team under budget 1000 for season 2021.}
\label{tab:best_2021_players}
\end{table}

\subsection{Formations}
\begin{figure}[h]
\captionsetup[subfloat]{labelformat=empty}
\centering
\setkeys{Gin}{width=0.48\linewidth, height=6cm} % common settings for images sizes 
\subfloat[]{\includegraphics[]{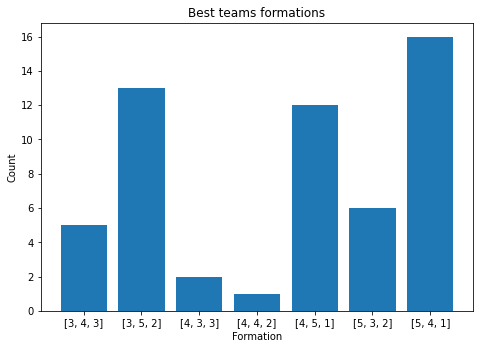}\hfill}
\subfloat[]{\includegraphics[]{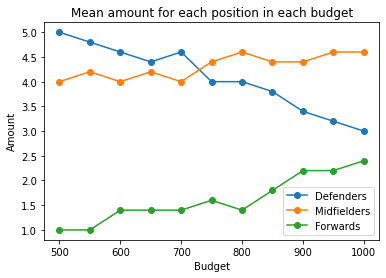}}
\caption{The left figure shows how many times each formation was the best formation for each season of FPL.  Count refers to the number of times the formation was the highest scoring within its budget class. The right figure shows the mean cost of each position subject to each budget constraint.  Amount refers to the number of players of each of the three types in the best team with the budget given on the horizontal axis. This reveals a trend that for a lower budget, it is preferable to spend more money on defenders and less on forwards, while for a higher budget the opposite is the case. }
\label{fig:bestformation}
\end{figure}

We also analysed which formation that gave the best result for each budget. The results can be seen in Figure  \ref{fig:bestformation}. We can see that three formations stood out, namely 3-5-2, 4-5-1 and 5-4-1. When taking the mean of the best position for each budget we can see an interesting trend, visualised in Figure \ref{fig:bestformation}: for a lower budget it is preferable to have more defenders and fewer forwards, while for a higher budget it is the opposite. 

\begin{figure}[h]
\captionsetup[subfloat]{labelformat=empty}
\centering
\setkeys{Gin}{width=0.44\linewidth,height=5cm} % common settings for images sizes 
\subfloat[]
{\includegraphics{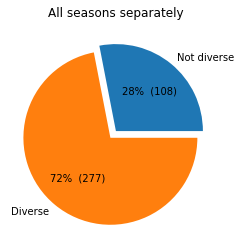}    
\hfill}
\subfloat[]
{\includegraphics{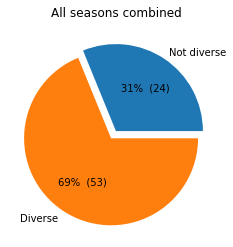}    
}
\caption{This shows the percentage of best teams in all budgets that have diverse or non diverse salary distributions.  On the left, we calculated the best teams for each season separately, for each budget constraint.  Of these, 72 \% have diverse salary distributions.  On the right, we calculated the best teams for each budget constraint, allowing players to be selected from any of the five seasons.  Of these, 69 \% have diverse salary distributions. }
\label{fig:pie}
\end{figure}

\subsection{Salary distributions within the best teams} \label{s:best_teams} 
We investigated the salary distributions of the best teams in every formation and budget. 
The total number of best teams is 
\[5*7*11=385\] 
because we assessed 5 seasons, 7 formations, and 11 budgets in FPL. Figure \ref{fig:pie} shows that 72\% of the best teams' salary distributions were considered diverse. The distributions of costs between the players were similar when we checked all seasons separately and when we combined the players from all seasons. Motivated by this fact, we analyzed teams that could contain players from different seasons.  In this analysis, 69\% of the best teams were considered to be diverse.   
Consequently, we observe that a diverse salary distribution is a predominant characteristic of the best preforming FPL teams.  

\begin{figure}[h]
\centering
%\vspace{1cm}	% Adjust vertical spacing here
\includegraphics[width=\linewidth]{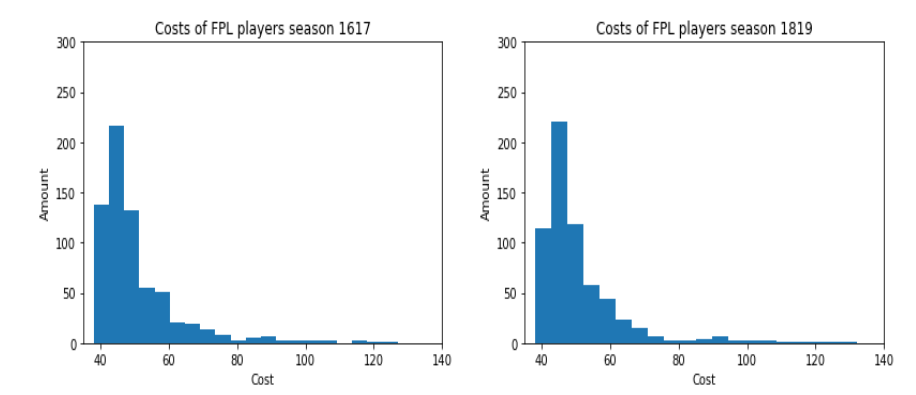}
\caption{This shows the distribution of salaries across all players.  There are approximately 700 players in each of the seasons.  Approximately 230 players have the same salaries, corresponding to about 33\% of all the players.}
\label{fig:all_salaries}
\end{figure}

We note that if a team were chosen completely randomly, then the probability that the salary distribution would be classified as diverse is quite low, because the majority of players have nearly identical salaries; see Figure \ref{fig:all_salaries}.  Approximately 33\% of the players have salaries in a single bin, so with a purely randomly chosen team, approximately 33\% of the team, or 4 players, would have salaries contained in a single bin.  Moreover, there are two additional salary bins of players with approximately 120 players per bin.  So, similarly, with a purely randomly chosen team, at least approximately 2 more players would also have salaries contained in a single bin. Consequently, a purely randomly chosen team is \em unlikely \em to have a diverse salary distribution, because 6 players would have salaries contained within just 2 bins.  Consequently, even if all the other 5 players' salaries were contained in different bins, there would still be 4 empty bins, resulting in a non-diverse salary distribution. 

\subsection{Distributions of other variables within the best teams} \label{s:diversity_vars} 
\begin{figure}[H]
\setkeys{Gin}{width=0.33\linewidth,height=4.4cm}
\centering
{\includegraphics{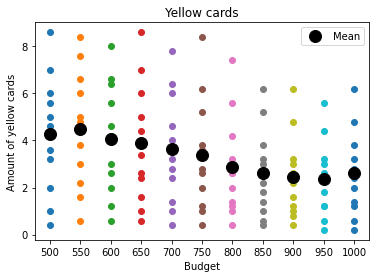}}\hfill
{\includegraphics{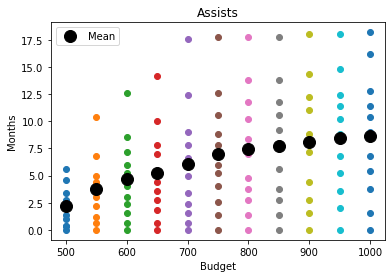}}\hfill 
{\includegraphics{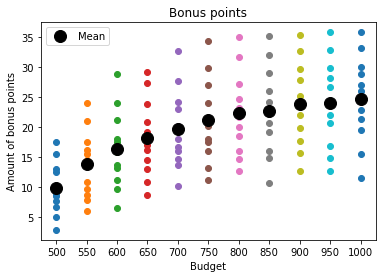}} 

{\includegraphics{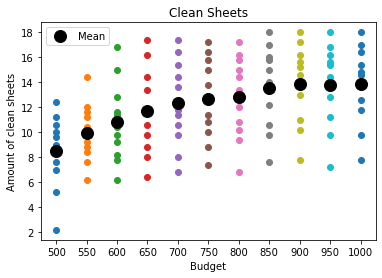}}\hfill  
{\includegraphics{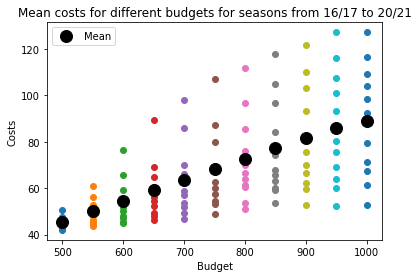}}\hfill
{\includegraphics{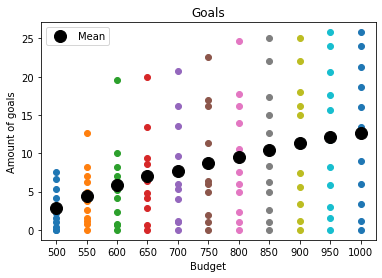}}\hfill 
  
{\includegraphics{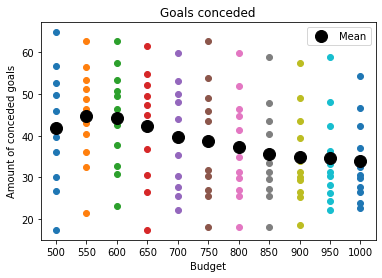}}\hfill
{\includegraphics{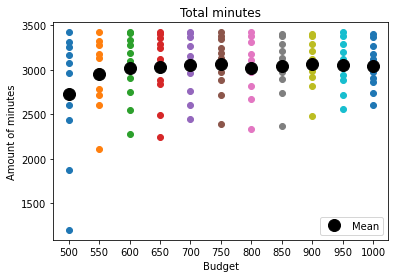}}
{\includegraphics{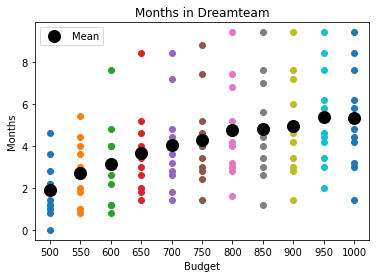}}\hfill

{\includegraphics{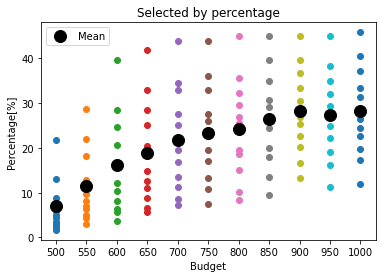}}\hfill
{\includegraphics{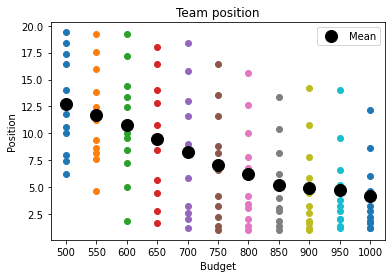}}\hfill
{\includegraphics{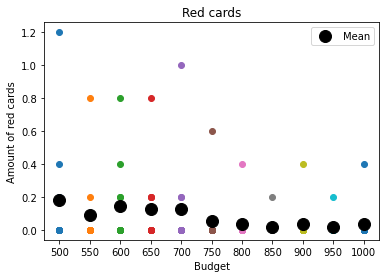}}\hfill
\caption{The distribution of different variables for the best teams in the FPL seasons.  The black dots represent the mean of each variable for each budget.  The colored dots show the values for each player sorted according to cost.}
\label{fig:other_var_div}
\end{figure}

A diverse salary distribution is a characteristic that is common to the majority of the most successful FPL teams as shown in \S \ref{s:best_teams}; see Figure \ref{fig:pie}.  As one can see from Figure \ref{fig:all_salaries}, a randomly chosen team is \em unlikely \em to have a diverse salary distribution. Is there diversity in \em other \em variables within the best teams?  We assessed this for several variables that are essential to football.  The results for the different variables are presented in Figure \ref{fig:other_var_div}.  Each colored dot represents the mean value for each sorted (according to cost) player.  The black dots represent the mean for each budget.  The variables we assessed are essential to football: 
\begin{itemize} 
\item Yellow cards. 
\item Assists. 
\item Bonus points: given to a player according to the rules of FPL. 
\item Clean sheets: if the team doesn't concede any goal.
\item Mean costs for different budgets for seasons 2017 to 2021.
\item Goals. 
\item Goals conceded. 
\item Total minutes:  all minutes the player played during the season. 
\item Months in Dreamteam: how many months the player was selected to the Dreamteam.
\item Selected by percentage:  how many percentage of managers that selected that player. 
\item Team position:  the placement of each players team in the league. 
\item Red cards.
\end{itemize} 
The distributions for 11 of these 12 variables are diverse.  The only variable with a non-diverse distribution is the number of red cards.    This may indicate that it is better to have more players that all contribute different amounts across a range of values, rather than having just a few star players carrying the team.   With the exception of red cards, there is a large spread within each variable and budget constraint, corresponding to diversity across these variables.  

%%%%%%%%%%%%%%%%%%%%%%%%%%

\section{Discussion} \label{s:discussion} 
Can one \em predict \em the fantasy team which would score the highest points in an upcoming football season?  Or, could one identify fantasy teams which are \em likely \em to be among the highest scoring in the next season?  Although we cannot answer these questions with the present study, we have made a fundamental step towards prediction by analyzing the characteristics of top performing teams with a range of budgets for the players' salaries.  Our results here show that diversity is a common feature of these top performing teams.  

There may be a more general underlying mechanism through which diversity amongst team members is beneficial to the team in competition with other teams. Indeed, this concept within high performing teams dates back to at least the 1980s; see for example the Belbin model \cite{belbin}.  There, Belbin considered different types of contributors in a team and showed that teams with these different types of contributors generally perform well.  Belbin's work is based on psychology and empirical studies.  More recently, Rowlett et. al. has formulated a purely theoretical mathematical approach for analyzing competing teams in \cite{diversity}. The \em Diversity Theorems \em in  \cite{diversity}, show that a team performs better if the individuals within it are diverse with regard to any competitive ability with a budget constraint on this competitive ability.  The Diversity Theorems are based on a game-theoretic model for competing teams comprised of individuals.  The individuals of a team each have a competitive ability that determines their success or defeat in competition with an individual from an opposing team.  Identical competitive abilities result in a tie, whereas different competitive abilities result in a win for the individual with the higher ability and a loss for the individual with the lower ability.  The team's success is calculated by amassing all cumulative wins and losses of the team's individuals.  If no constraint is imposed on the competitive ability values, then to achieve the strongest possible team, one would simply let the competitive ability values tend to infinity.  However, if a constraint is imposed, then certain distributions of competitive abilities amongst individuals outperform others.  The Diversity Theorems identify the best way to distribute competitive abilities subject to a constraint on the mean competitive ability.  There, a specific way of distributing competitive abilities is called a strategy.  The Diversity Theorems show that the best strategies are those that assign individuals a diverse range of competitive abilities \cite{diversity}.  It is interesting to note that although the scientific approaches of \cite{belbin} and \cite{diversity} are completely different, the results sound quite similar.   

One of major the challenges of utilizing the Diversity Theorems is to interpret the real-world meaning of `competitive ability.' This is one of the motivations for our investigation here of professional sports, because there is a correlation between an individual player's salary and their performance \cite{Yaldo2017}.  Identifying a player's salary with their competitive ability allows us to investigate the theoretical predictions.  Moreover, the `mean competitive ability' (mca) constraint of \cite{diversity} is equivalent to having a budget constraint for the salaries of the team.  This is a natural and realistic constraint, because every real-life team has a limited budget from which to pay the players' salaries.  We note, however that `competitive ability' is a mathematical concept, so it could be used to describe \em any \em trait.  Consequently, we analyzed not only the salary distributions but also the distributions across several other variables that are essential to football as shown in Figure \ref{fig:other_var_div}.  With both the salary distribution as well as the distributions across all variables (except red cards) demonstrated in Figures \ref{fig:pie} and \ref{fig:other_var_div}, respectively, the diversity observed across all these different variables shows that, in general, diversity is a prevalent characteristic of the best performing teams.  This echoes the theoretical predictions of the Diversity Theorems \cite{diversity} and the Belbin's empirical studies \cite{belbin}. 

Motivated by the results of this paper, further research should focus on generalizing the performed analyses to other contexts: Is diversity key only for fantasy football?  What about other teams? Both the Belbin model \cite{belbin} and the Diversity Theorems \cite{diversity} indicate the relevance of diversity.  In this paper we managed to transfer these theoretical truths to practice. However, this is only the first step. While our analyses can be applied to almost all teams where we have a constraint (such as the budget) and a measurable performance (such as scores), the results need to be checked for other contexts. For example, software projects require a diverse knowledge distributed among the team members. A couple of years ago, the idea of so-called cross-functional teams that have almost all required knowledge in the team, emerged in research and practice \cite{agilemanifesto}. Having all required knowledge in the team is also an aspect of diversity. Having only members with a similar set of skills and knowledge would be less beneficial. However, so far, there is no approach showing how to form a team given a number of possible team members. Transferring the results presented in this paper to such a context would not only influence the work of HR managers by helping to determine which skills the new team member needs in order to increase the team's diversity, but also affect the collaboration and the knowledge sharing in the teams. 

In this direction, it would also be interesting to analyze how diversity changes over time. That is, if one starts with a diverse team, how do the characteristics of players or team members influence the performance of others within the team? Does diversity decrease over time as, for example, knowledge or techniques are shared with the other team members? In the context of hypothetical teams (as we have in fantasy football) this line of thought cannot be analyzed, because these team members are not actually on the same team. Therefore, transferring this idea to real teams would be interesting.  We propose that further research should focus on strengthening the results obtained here and analyzing them with regard to the applicability in other contexts such as HR.

%\bibliographystyle{nar}
%\bibliography{ff_bib}

\begin{thebibliography}{10}

\bibitem{Adhikari2020}
A.~Adhikari, A.~Majumdar, G.~Gupta, and A.~Bisi, \emph{An innovative
  super-efficiency data envelopment analysis, semi-variance, and
  shannon-entropy-based methodology for player selection: evidence from
  cricket}, Annals of Operations Research \textbf{284} (2020), no.~1.

\bibitem{agilemanifesto}
K.~Beck, M.~Beedle, Ar. Van~Bennekum, A.~Cockburn, W.~Cunningham, M.~Fowler,
  J.~Grenning, J.~Highsmith, A.~Hunt, R.~Jeffries, et~al., \emph{The agile
  manifesto}, 2001.

\bibitem{belbin}
R.~M. Belbin, \emph{Management teams}, Heinemann, London, 1981.

\bibitem{BillingsFantasy2013}
A.~C. Billings and B.~J. Ruihley, \emph{The fantasy sport industry: Games
  within games}, Routledge, Taylor and Francis Group, 2013.

\bibitem{Brettenny2012}
W.~J. Brettenny, D.~G. Friskin, J.~W. Gonsalves, and G.~D. Sharp, \emph{A
  multi-stage integer programming approach to fantasy team selection: A
  twenty20 cricket study}, South African Journal for Research in Sport,
  Physical Education and Recreation \textbf{34} (2012), no.~1.

\bibitem{Davis2006}
N.~W. Davis and M.~C. Duncan, \emph{Sports knowledge is power: Reinforcing
  masculine privilege through fantasy sport league participation}, Journal of
  Sport and Social Issues \textbf{30} (2006), no.~3.

\bibitem{Dyreson2019}
M.~Dyreson, \emph{Looking backward and forward from the 24-million-word mark: A
  managing editor's perspective on the international journal of the history of
  sport in transition}, International Journal of the History of Sport
  \textbf{36} (2019).

\bibitem{Foster2020}
G.~Foster, N.~O'Reilly, and A.~D{\'a}vila, \emph{Sports business management},
  Routledge, Taylor and Francis Group, 2020.

\bibitem{gamson1976}
W.~A. Gamson, \emph{The strategy of social protest}, Dorsey Press, University
  of Michigan, 1975.

\bibitem{mastersthesis}
J.~Gullholm and J.~St{\aa}lberg, \emph{What makes a winning fantasy football
  team?}, Master's thesis, Chalmers Tekniska H\"ogskola and G\"oteborgs
  Universitet, 2022.

\bibitem{Holt2011}
R.~Holt, \emph{Sport and the british: A modern history}, Clarendon Press,
  Oxford, 2011.

\bibitem{Karthik2021}
K.~Karthik, Gokul~S. Krishnan, Shashank Shetty, Sanjay~S. Bankapur, Ranjit~P.
  Kolkar, T.~S. Ashwin, and Manjunath~K. Vanahalli, \emph{Analysis and
  prediction of fantasy cricket contest winners using machine learning
  techniques}, Advances in Intelligent Systems and Computing, vol. 1176, 2021.

\bibitem{Kaur2020}
G.~Kaur and G.~Jagdev, \emph{Analyzing and exploring the impact of big data
  analytics in sports science}, Indo - Taiwan 2nd International Conference on
  Computing, Analytics and Networks, Indo-Taiwan ICAN 2020 - Proceedings, 2020.

\bibitem{King2017}
N.~King and A.~LeBoulluec, \emph{Projecting a quarterback's fantasy football
  point output for daily fantasy sports using statistical models},
  International Journal of Computer Applications \textbf{164} (2017), no.~4.

\bibitem{Kissel2017}
R.~Kissell and J.~Poserina, \emph{Optimal sports math, statistics, and
  fantasy}, Academic Press, Elsevier, 2017.

\bibitem{Rodolfo1892}
Rodolfo Lanciani, \emph{Gambling and cheating in ancient rome}, The North
  American Review \textbf{155} (1892), no.~428, 97--105.

\bibitem{Naha2021}
S.~Naha, \emph{Flight of fantasy or reflections of passion? knowledge, skill
  and fantasy cricket}, Sport in Society \textbf{24} (2021), no.~8.

\bibitem{Ploeg2021}
A.~J. Ploeg, \emph{A new form of fandom: How free agency brought about
  rotisserie league baseball}, International Journal of the History of Sport
  \textbf{38} (2021), no.~1.

\bibitem{diversity}
J.~Rowlett, C.J. Karlsson, and M.~Nursultanov, \emph{Diversity strengthens
  competing teams}, Royal Society Open Science \textbf{9} (2022).

\bibitem{Ruihley2021}
B.~J. Ruihley, A.~C. Billings, and N.~Buzzelli, \emph{A swiftly changing tide:
  Fantasy sport, gambling, and alternative forms of participation}, Games and
  Culture \textbf{16} (2021), no.~6.

\bibitem{RuihleyVoice2021}
B.~J. Ruihley and J.~Chamberlin, \emph{The history and evolution of the fantasy
  sport voice: An oral account of the major aspects forming the fantasy sports
  and gaming association}, International Journal of the History of Sport
  \textbf{38} (2021), no.~1.

\bibitem{Forbes}
D.~Saul, \emph{Number of people betting on sports doubled in 2021, poll finds,
  as new york reports an explosive debut}, 2022,
  \href{https://www.forbes.com/sites/dereksaul/2022/01/19/number-of-people-betting-on-sports-doubled-in-2021-poll-finds-as-new-yorks-reports-an-explosive-debut/?sh=7c02019d5e7d}{Forbes
  Magazine}.

\bibitem{South2019}
C.~South, R.~Elmore, A.~Clarage, R.~Sickorez, and J.~Cao, \emph{A starting
  point for navigating the world of daily fantasy basketball}, American
  Statistician \textbf{73} (2019), no.~2.

\bibitem{Thuillier2004}
J.~P. Thuillier, \emph{Le sport dans la civilisation {\'e}trusque : entre
  gr{\`e}ce et rome}, {\'E}tudes balkaniques \textbf{11} (2004), 13--32.

\bibitem{Git}
A.~Vaastav, \emph{Fantasy-premier-league}, 2013,
  \href{https://github.com/vaastav/Fantasy-Premier-League}{GitHub repository}.

\bibitem{Yaldo2017}
L.~Yaldo and L.~Shamir, \emph{Computational estimation of football player
  wages}, International Journal of Computer Science in Sport \textbf{16}
  (2017).

\end{thebibliography}
%\bibliographystyle{amsplain}

\end{document}